\documentclass{article}

\usepackage{arxiv}
\usepackage[textwidth=8em,textsize=small]{todonotes}
\usepackage{amsmath}
\usepackage{natbib}
\usepackage{booktabs}
\usepackage{hyperref}
\usepackage{dsfont}
\usepackage{multirow}
\usepackage{tabularx}
\usepackage{setspace}
\usepackage{lineno}
\usepackage{float}

\title{Spatio-temporal Downscaling Emulator for Regional Climate Models}

\author{
	Luis A. Barboza \\
	Centro de Investigaci\'on en Matem\'atica Pura y Aplicada-Escuela de Matem\'atica\\
	Universidad de Costa Rica,\\
	San Jos\'e, Costa Rica \\
	\texttt{luisalberto.barboza@ucr.ac.cr} \\
	\And
	Shu Wei Chou Chen \\
	Centro de Investigaci\'on en Matem\'atica Pura y Aplicada-Escuela de Estad\'istica\\
	Universidad de Costa Rica,\\
	San Jos\'e, Costa Rica \\
	\texttt{shuwei.chou@ucr.ac.cr} \\
	\And
	Marcela Alfaro C\'{o}rdoba \\
	Department of Statistics\\
	University of California, Santa Cruz\\
	Santa Cruz, CA, USA \\
	\texttt{macordob@ucsc.edu} \\
	\And
	Eric J. Alfaro \\
	Centro de Investigaciones Geof\'isicas, Escuela de F\'isica and Centro de Investigaci\'on en Ciencias del Mar y Limnolog\'ia\\
	Universidad de Costa Rica,\\
	San Jos\'e, Costa Rica \\
	\texttt{erick.alfaro@ucr.ac.cr} \\
	\And
	Hugo G. Hidalgo \\
	Centro de Investigaciones Geof\'isicas and Escuela de F\'isica\\
	Universidad de Costa Rica,\\
	San Jos\'e, Costa Rica \\
	\texttt{hugo.hidalgo@ucr.ac.cr} \\
}

\begin{document}
	\onehalfspacing
	
	\maketitle
	\begin{abstract}
 
	Regional Climate Models (RCM) describe the meso scale global atmospheric  and  oceanic  dynamics and serve as dynamical downscaling models. In other words, RCMs use atmospheric and oceanic climate output from General Circulation Models (GCM) to develop a higher  resolution climate output. They are computationally demanding and, depending on the application, require several orders of magnitude of computer time more than statistical climate downscaling. In this paper we describe how to use a spatio-temporal statistical model with varying coefficients (VC), as a downscaling emulator for a RCM using varying coefficients. In order to estimate the proposed model, two options are compared: INLA, and varycoef. We set up a simulation to compare the performance of both methods for building a statistical downscaling emulator for RCM, and then show that the emulator works properly for NARCCAP data. The results show that the model is able to estimate non-stationary marginal effects, which means that the downscaling output can vary over space. Furthermore, the model has flexibility to estimate the mean of any variable in space and time, and has good prediction results. INLA was the fastest method for all the cases, and the approximation with best accuracy to estimate the different parameters from the model and the posterior distribution of the response variable. 
	\end{abstract}

\section{Introduction}\label{intro}

Regional Climate Models (RCM) describe the meso scale global  atmospheric  and  oceanic  dynamics and serve for downscaling coarser resolution climate models. In other words, RCMs use atmospheric and oceanic climate output in General Circulation Models (GCM) to develop a higher resolution  climate output,  which  is  influenced  by  smaller scale climate dynamics and a higher resolution topography, coastlines, inland water or land-surface characteristics. They represent a tool for assessing climate variability and change impacts,  to develop  seasonal  climate  predictions,  and  are a  powerful  tool  in general for  improving  our  understanding  of regional climate processes \citep{wang04,Amador2009}.  

The  North  American  Regional  Climate  Change  Assessment Program (NARCCAP) is an international program that has as a main goal generating climate scenarios for use in impacts research \cite{Mearns2009}. They have proposed and applied a dynamical downscaling technique that consists of  embedding  RCMs  within  GCMs  to  obtain  higher  resolution evaluations of the climate model over the domain of interest, using different combinations of RCMs and GCMs. As \cite{Wilby1997} state: ``RCMs are computationally demanding and require orders of magnitude more computer time than statistical downscaling to compute equivalent scenarios''. Since this dynamical downscaling technique is computationally  expensive,  a downscaling emulator can serve as a fast statistical approximation to  perform sensitivity analysis, to downscale a large set of GCM data  or  for  making  informed  prior judgments relating the GCM drivers for an RCM; therefore is of high value. 

Previous work on statistical emulators from \cite{Castruccio2014}, \cite{overstall2016multivariate}, \cite{Mearns1999}, \cite{OHagan2006} and \cite{Hernanz2022} are concentrated on climate model's projections and on computer experiment output. In this work, we describe how to use a spatio-temporal statistical model with varying coefficients (VC), as a downscaling emulator for a RCM. The general structure of our proposed emulator follows the lines of the wide field of surrogate modeling, according to \cite{Sacks1989,Gramacy2012}, in the sense that we incorporate zero-mean gaussian processes as modeling ingredients. Other experiments to emulate regional models using the NARCCAP output and downscaling techniques have been performed, as described in \cite{Wood2004} and \cite{Laflamme2016}. Nevertheless, to the best of our knowledge there are no attempts to emulate RCM using VC models. Despite of the above, other fields have employed VC models to solve research problems, for example in epidemiology \cite{Wang2022} and ecology \cite{Finley2011}.

Most of the models that use VCs are based on the seminal work by \cite{Gelfand2003a}. A myriad of applications for this model have been developed in epidemiology, ecology, among others, and some in climate models, such as \cite{liying2019}, but not for downscaling as in this study. These models allow for marginal effects to be non-stationary over space, time or both, and thus offer a higher degree of flexibility than a simple fixed coefficient spatio-temporal model. Due to the computational burden that inverting big matrices including in these models can have with big data, a proposal that uses VCs in space, time or both, needs to be computationally efficient. 

Comparisons between methods for analyzing large spatial data, such as the one presented in \cite{Heaton2018}, have compared the speed and accuracy of spatial models but have not included VC methods. In this work, two methods for estimating spatio-temporal downscaling emulators that use VC models are being compared: i) Integrated Nested Laplace Approximation (INLA), as presented in \citep{Rue2009}; and ii) the approach from \cite{Dambon2021} (varycoef). Both are implemented in R packages that are already published: \textit{INLA} and \textit{varycoef}, respectively. However, varycoef is implemented using frequentist approach while we adapted \textit{varycoef} to the Bayesian approach.

A Multi-resolution (MRA) approach such as the one in \cite{katzfuss_multi-resolution_2017} can help with the computational burden, and it has been shown to work with massive spatially distributed data \cite{Katzfuss2016} in the context of regular Gaussian processes. However, its implementation is not straight forward in R, since there are no packages on CRAN for that. In this work, we concentrate on showing the results for INLA and compare them with the method of \cite{Dambon2021} instead. To the best of our knowledge, INLA has limited work on VC models, but \cite{Blangiardo2013} have shown how to work with linear models with random effects. Additionally, the method of \cite{Dambon2021} is the only method with a R package that has been exclusively proposed for spatially VC models to use on medium-size data sets. The objective of this paper is to compare the performance of INLA and varycoef for building a statistical downscaling emulator for RCM, and then show that the emulator works properly for NARCCAP data.

The outline of this article is as follows: Section \ref{sec2} describes the spatio-temporal downscaling emulator, and gives an overview on the estimation details for each method. Section \ref{sec3} describes two simulation studies: the first one investigates the performance of each of the methods when estimating the statistical emulator, and the second one explores the fitting of two different models to emulate data with similar characteristics as the real data application. Section \ref{sec4} presents the NARCCAP data, the motivation for the problem, and the results from a real data application of the model. Finally, Section \ref{sec5} presents the discussion and conclusions. 

\section{Statistical Methods}\label{sec2}

\subsection{Spatio-temporal Downscaling Emulator}

\begin{figure}[htp!]
	\centering
	\includegraphics*[scale=0.35]{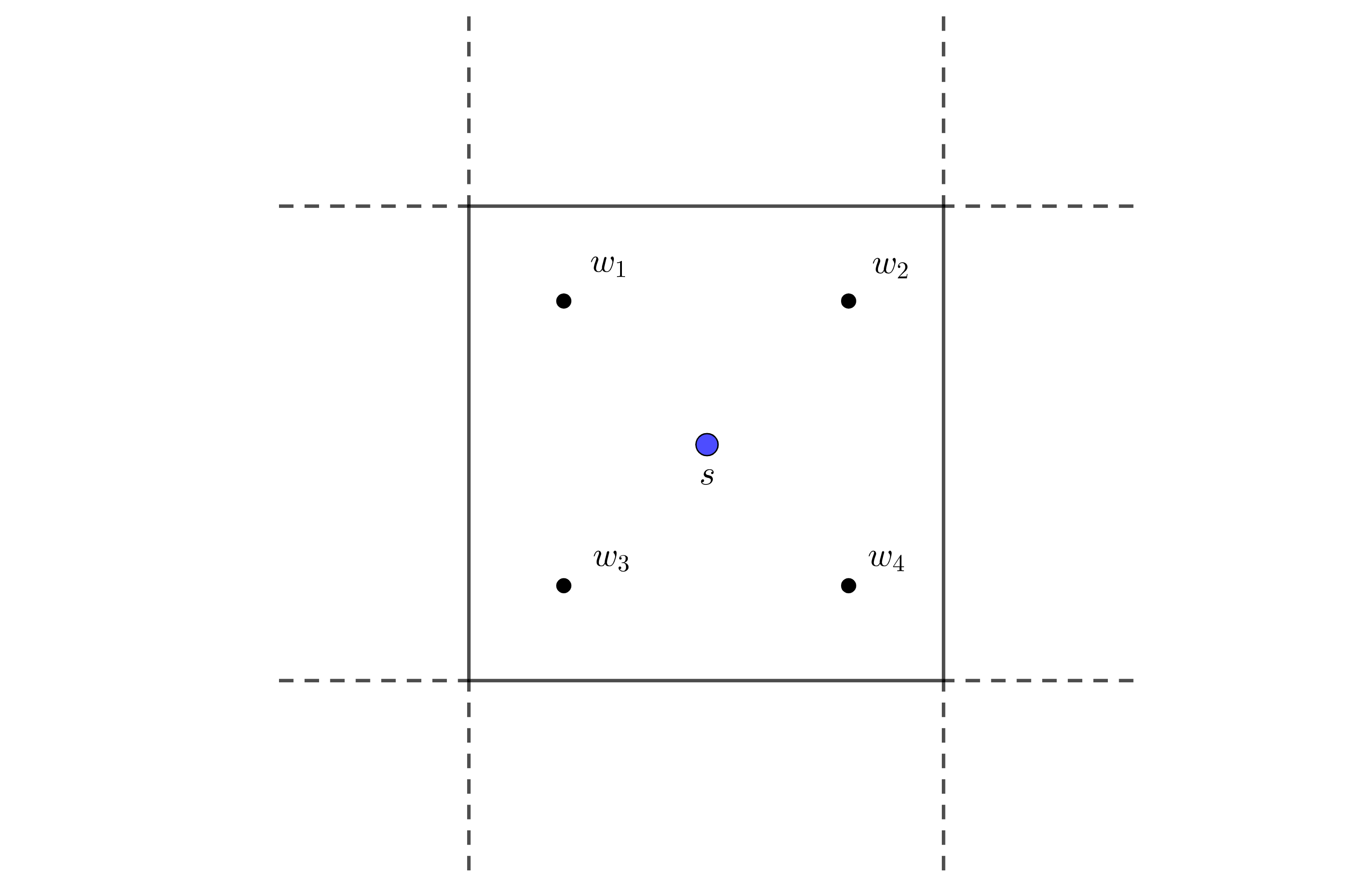}
	\caption{Illustration of points in $\mathcal{S}$ (blue) and $\mathcal{W}$ (black).}
	\label{diagrama}
\end{figure}

Consider a climate-related variable represented as a random variable $C_t(s)$ indexed by time $t$ and spatial location $s\in \mathcal S$. The spatial set $\mathcal S$ is assumed to be regularly-spaced with a fixed resolution. Assume there exists a set of $q$ covariates linearly related with the dependent variable $C_t(s)$ in the following way:
\begin{align}\label{Global_eq}
	C_t(s)=\alpha+\beta^\top X_t(s)+\epsilon_t(s),
\end{align} 
where $\alpha$ and $\beta$ are random parameters and $\epsilon_t(s)$ is a white noise in both space and time with variance $\zeta^2$. The main idea is to model the random field $C_t(\cdot)$ over a set of finer resolution on the same spatial domain $\mathcal S$. Let $\mathcal W$ be such set, and assume that the resolution satisfies that each location $s\in \mathcal S$ can be placed as a center of a rectangle containing several (or none) locations $w_1,\ldots,w_K$ on the finer set $\mathcal W$. Figure \ref{diagrama} shows how the location of both sets can be visualized when $K=4$.

Consider an arbitrary point $w\in \mathcal{W}$ which is located within the rectangle whose center is $s$. We assume that \eqref{Global_eq} holds on this scale, but we adjust the local relationship among the response and predictor as follows:
\begin{align}\label{Regional_eq}
	C_t(w)=[\alpha+\alpha_t^r(w)]+[\beta+\beta_t^r(w)]^\top X_t(s)+\epsilon_t(s)+\gamma_t(w),
\end{align}
where 
\begin{itemize}
	\item $\alpha_{\cdot}^r(\cdot)\sim N(\beta_0,\Sigma_0(\theta_0))$,
	\item $\beta_{\cdot}^r(\cdot)\sim N(\beta_1,\Sigma_1(\theta_1))$,
	\item $\gamma_t(\cdot)\stackrel{i.i.d}{\sim} N(0,\tau^2)$
\end{itemize}
and $\Sigma_0$ and $\Sigma_1$ are separable spatio-temporal covariance matrices parameterized with the vectors $\theta_0$ and $\theta_1$ respectively and the parameters $\beta_0$ and $\beta_1$ take into account any resulting bias due to the resolution change. The random terms $\alpha_t^r(w)$, $\beta_t^r(w)$ and $\gamma_t(w)$ account for the increase in uncertainty due to the downscaling process. Based on the above, we are interested in the following model:
\begin{align}
	Y_t(w):=C_t(w)-C_t(s)=\alpha_t^r(w)+\beta_t^r(w)^\top X_t(s)+\gamma_t(w),
	\label{modelo}
\end{align}
where $C_t(s)$ is treated as a known covariate in this model and hence we are interested in the modeling of the difference $Y_t(w)$ in order to get predictions on the finer variable $C_t(w)$. Note that $\alpha_{\cdot}^r(\cdot)$ and $\beta_{\cdot}^r(\cdot)$ represent random spatio-temporal correction terms on $\alpha$ and $\beta$ in \eqref{Global_eq} due to the use of the predictor $X_t(s)$ over the coarser set $\mathcal S$ and \eqref{Regional_eq} the change in scale between the sets $\mathcal S$ and $\mathcal W$.  

Under a Bayesian approach, we are interested in the conditional distribution of the dependent vector $\mathbf Y:=(Y_t(w))_{t,w}$ given the parameters $\Phi:=(\beta_0,\beta_1,\theta_0,\theta_1, \tau^2)$:
\begin{align*}
	\mathbf Y|\Phi \sim N(\beta_0+\mathbf X^\top \beta_1,\Sigma_Y),
\end{align*}
where 
\begin{align*}
	\Sigma_{\mathbf Y}=\mathbf X^\top  \left(\Sigma_0(\theta_0)+\Sigma_1(\theta_1)\right) \mathbf X+\tau^2 I_n
\end{align*}
and $\mathbf X$ is a block-diagonal matrix whose $i$-th sub-matrix is $X^\top(s_i)$, $i=1,\ldots,n$ and $n$ is the number of locations in $\mathcal S$. In order to evaluate the likelihood of $\mathbf Y$ we need to evaluate the inverse and determinant of $\Sigma_{\mathbf Y}$ which can be time-consuming for a combination of a large $n$ and/or a large number of times $t$. Several methods have been proposed to bypass calculating the inverse and determinant of $\Sigma_{\mathbf Y}$ for large data using approximations. In the next section, these methods are explained in light of our model.

\subsection{Approximation Methods}

Depending on both the temporal and spatial dimensions, the inverse of the resulting covariance matrix in its Gaussian likelihood can be time-consuming to compute, specially when any (or both) dimensions are large. The likelihood calculation is required to estimate the parameters of \eqref{modelo}, either by a frequentist or Bayesian approach. For the purpose of this work, we are interested in the Bayesian inference of the model because we want to measure the uncertainty due to downscaling in a more exact way, compared with asymptotic alternatives provided by classical methods. We aim to remedy the computational issues by means of varycoef and INLA because those methods are modern alternatives to accelerate the direct computation of the likelihood function. 

\subsubsection{Dambon Approach}

As we mentioned before, the model in \eqref{modelo} is a spatially-varying model when the time $t$ is considered as fixed. A first approach that we use in this case is the likelihood approximation according to \cite{Dambon2021}. This approach has two important assumptions, first it assumes mutual prior independence of the random effects $\alpha^r(\cdot)$ and $\beta^r(\cdot)$ and furthermore it assumes that each random effect has tapering on its covariance matrix. The first assumption allows the covariance matrix of $Y$ in model \eqref{modelo} to be written as:

\begin{align*}
	\Sigma_Y=\sum_{i=1}^p \Sigma^{j}  \odot \mathds{X}^{j}+\tau^2 I_n,
\end{align*} 
where $\odot$ is the Hadamard product and $\mathds{X}^{j}:= x^{(j)}\left(x^{(j)}\right)^T$ uses the $j$th covariate $x^{(j)}$. In order to ease the computation burden, \cite{Dambon2021} incorporate the tapered covariance matrix, $\Sigma_{tap}^{j}:=\Sigma^{j} \odot C_{\rho^*}$, where $\rho^*$ is the tapering range. In this case, it is a sparse matrix with $[\Sigma_{tap}^{j}]_{kl}=0$ for $||s_k-s_l||\geq \rho^*$. As a result, the covariance matrix can be simplified as: 

\begin{align*}
	\Sigma_{Y,tap}=\sum_{i=1}^p \Sigma_{tap}^{j}  \odot \mathds{X}^{j}+\tau^2 I_n.
\end{align*} 

The tapering, together with the sparsity of the matrix $\textbf{X}$ let the execution time to be reduced. This approach is fully implemented with maximum likelihood estimation on the R package varycoef, available on CRAN. We adapted the approach by using MCMC estimation. 

\subsubsection{Integrated Nested Laplace Approximation}

The computational challenge of MCMC inference in model \eqref{modelo} is due to the frequent evaluation of the likelihood during the Metropolis-Hastings steps. A possible strategy is to rely on Integrated Nested Laplace Approximation (INLA) to accelerate the MCMC procedure. The INLA approach is applicable to a general specification for which the mean $\eta_i$ of the observations $y_i$ follows a linear structure:
\begin{align}\label{eq:meanINLA}
	\eta_i = \alpha +\sum_{m=1}^M\beta_mx_{mi}+\sum_{l=1}^Lf_l(z_{li}),
\end{align}
where $\alpha$ represents an intercept, the coefficients
$\mathbf{\beta} = (\beta_1,\ldots,\beta_M)$ relate $M$ covariates
$(x_1,\ldots,x_M)$ to $\eta_i$, and $f = \{f_1(\cdot),\ldots,f_L(\cdot)\}$ is a collection of
random effects defined on a set of $L$ covariates $(z_1,\ldots,z_L)$ (see
\citealp{Rue2009} and \citealp{Blangiardo2013}). 

In model \eqref{modelo}, if we consider $y_i=\eta_i=Y_t(w)$ and the varying intercept and slope are the latent variables with fixed effects as their respective means $\beta_0$ and $\beta_1$, then our models fall into the general specification of INLA. The main objective of our Bayesian estimation is to compute the
marginal posterior distribution of each parameter in $\Phi$. To attain computational advantages, INLA assumes that the prior of
vector $\Phi$ is a multivariate normal random vector with a precision matrix that depends on a certain set of hyperparameters. INLA further 
approximates the conditional distribution of $\Phi$ by using a simplified Laplace Approximation together with optimization procedures (see \citealp{Rue2009}). 

\section{Simulation Study}\label{sec3}

Two independent simulation studies were performed to evaluate the predictive ability of the proposed emulator. A first exercise was done to compare INLA (\citealp{Rue2009}) and varycoef (\cite{Dambon2021}, R Package) in a spatial temporal setting. This was done to choose the final estimation method for the emulator. The results from this simulation are presented in Appendix \ref{AppendixA}. In summary, INLA outperforms varycoef in predictive capacity and elapsed time. Furthermore, INLA is a better alternative in more-complex settings as the spatio-temporal one presented in Appendix \ref{AppendixA}.

A second simulation study was done with the objective of evaluating the performance of the downscalling approximation using INLA. Nine independent settings to generate data were used with a spatio-temporal varying intercept following a random noise process (see \eqref{Regional_eq}), where the spatial covariance structure. Three different variability scenarios were combined and three different spatial resolutions. The same model used to generate the data was fitted using INLA, and assuming a) independence in the errors, and b) assuming an AR(1) structure with $\rho=0.8$. 

We study the in-sample and out-of-sample fitting performance of the same model, and the different scenarios to generate the data. Our main focus is to describe how well we can estimate a simulated response according to \eqref{Regional_eq}, and we compare goodness-of-fit performance for different structures of spatial correlation when estimating the response variable and the parameters involved, according to elapsed time and the final model estimation in \eqref{modelo}. For that, we use the following statistics: mean elapsed time to estimate the model \eqref{modelo} (in minutes), Mean Squared Error (MSE) and Interval Score (IS) \citep{Gneiting2007a} and for each estimated model. The latter metric is defined as:

\begin{align}
    \label{eq:IS}
	IS_{\alpha}=\frac 1 n \sum_{t,w}\left[(U_t(w)-L_t(w))+\frac{2}{1-\alpha}(L_t(w)-Y_t(w))\cdot 1_{Y_t(w)<L_t(w)}\right. \\
	  \left.+\frac{2}{1-\alpha}(Y_t(w)-U_t(w))\cdot 1_{Y_t(w)>U_t(w)}\right],
\end{align}
where $n$ is the number of observations, $Y_t(w)$ is the  response, $L_t(w)$ is the $(1-\alpha)$\%-predictive lower bound and $U_t(w)$ is the $(1-\alpha)$\%-predictive upper bound at location $w$ and time $t$ over the period of study. $IS_{\alpha}$ can be interpreted as an utility function in interval estimation that addresses width as well as coverage. This means that estimations with narrow prediction intervals are rewarded, and whenever the observation is outside of the interval, there is a penalty of size $\alpha$. Previous studies employed this metric to evaluate the predictive capacity of different models on a diversity of environmental studies (see \cite{Barboza2014,Barboza2019,Bracher2021,Gao2022}).

In the spatial region $\left[0,20\right]^2$, we set three different resolutions for the regularly-spaced set $\mathcal S$ and the finer set $\mathcal W$. If $(g \times g)$ and $(r \times r)$ denote the resolution size for $\mathcal S$ and $\mathcal W$ respectively, we will denote ($g \times r$) as a way to simplify the notation for a single scenario. To be precise, we use $(20 \times 10)$ as Resolution 1, $(40 \times 20)$ as Resolution 2, and $(60 \times 10)$ as Resolution 3. Moreover, $T=12$ periods of time were simulated for each scenario. For each simulated data, we divided it into two sets: a training data set containing the first $5/6$ periods of time, and a testing data set, containing the last $1/6$ periods of time. For the three different variability scenarios, we set the spatial varying intercept $\alpha_r(\omega)$ with spatial covariance structure Mat\'ern$\theta_0$ with $\theta_0=(\phi=5,\sigma ,\nu=1)$, where $\sigma= 0.003,0.0003,0.00003$. Recall that $\phi$ and $\sigma$ are known as the range and variance parameters of the Mat\'ern covariance structure and $\nu$ is its smoothness parameter. For each of the nine scenarios we generate 30 replications.

At these locations, we simulate two models according to \eqref{Global_eq} and \eqref{Regional_eq} with $q=0$, and $\alpha=5.707$, $\beta_0=5.706$, $\epsilon_t(s)\stackrel{i.i.d}{\sim} N(0,\zeta^2)$ with $\zeta^2=0.001$ and $\tau^2=1/700000$. We choose these parameter values based on the calculated values for the climate variable of interest in Section \ref{sec4}. Furthermore, in order to recover the climate variable and compute its predictive metrics, we transformed back the simulated log-transformed ${C_t(\cdot)}$ using an exponential function (for more details, see Section \ref{sec4}).

\subsubsection*{Simulation Results}

Data was analyzed using a penalized complexity (PC) prior in all cases according to \cite{Franco2019}. These priors avoid overfitting, and are implemented in the INLA R package. Table~\ref{table_goodnessfit} presents the goodness of fit metrics by model and resolution, and Table~\ref{table_time} includes the elapsed time in minutes for each of the resolutions and models.

\begin{table}
	\caption{\label{table_goodnessfit} Prediction metrics by model, scenario and resolution}
	
	\centering
	\begin{tabular}[t]{cclrrrrrr}
		\hline
		\multirow{3}{*}{Model} & \multirow{3}{*}{Scenario} & \multirow{3}{*}{Data} & \multicolumn{6}{c}{Resolution} \\
		\cline{4-9}
		\multicolumn{1}{c}{} & \multicolumn{2}{c}{} & \multicolumn{2}{c}{1} & \multicolumn{2}{c}{2} & \multicolumn{2}{c}{3} \\
		\cline{4-5} \cline{6-7} \cline{8-9}
		&  &  & MSE & $IS_{.95}$ & MSE & $IS_{.95}$ & MSE & $IS_{.95}$ \\
		\midrule
		\multirow{6}{*}{1} & \multirow{2}{*}{1} & Training & 271.7244 & 495.0508 & 272.8882 & 501.3870 & 272.8240 & 499.5025\\
	&  & Testing & 272.6720 & 496.3769 & 273.6045 & 501.3925 & 272.1695 & 498.8784\\
		& \multirow{2}{*}{2} & Training & 26.9863 & 142.8168 & 27.1295 & 145.6285 & 27.1019 & 144.4780\\
		&  & Testing & 27.2869 & 143.6768 & 27.3409 & 145.8234 & 27.2076 & 144.2540\\
		& \multirow{2}{*}{3} & Training & 2.6793 & 34.4497 & 2.7136 & 35.8174 & 2.7252 & 36.6192\\
		&  & Testing & 2.8473 & 35.5102 & 2.8477 & 36.1113 & 2.8371 & 36.1891\\
		\midrule
		\multirow{6}{*}{2} & \multirow{2}{*}{1} & Training & 270.6382 & 496.5986 & 274.6296 & 505.4288 & 273.2237 & 505.3798\\
		&  & Testing & 273.8565 & 497.7994 & 274.5304 & 505.5153 & 272.7651 & 505.4381\\
		& \multirow{2}{*}{2} & Training & 26.9412 & 145.4542 & 27.3155 & 147.1426 & 27.1747 & 147.0658\\
		&  & Testing & 27.3255 & 146.0270 & 27.4488 & 147.5109 & 27.2652 & 147.2583\\
		& \multirow{2}{*}{3} & Training & 2.6833 & 35.0120 & 2.7314 & 35.9817 & 2.7268 & 36.6530\\
		&  & Testing & 2.8372 & 35.9748 & 2.8640 & 36.3788 & 2.8393 & 36.2942\\
		\bottomrule
	\end{tabular}
\end{table}

From Table~\ref{table_goodnessfit}, it is important to note that the results per model are not significantly different in terms of MSE and IS, but model 1 has slightly smaller IS than model 2, particularly as the resolution increases. When comparing between resolutions, there is no specific trend when the resolution increases or decreases, and the difference between 1, 2, and 3 are relatively small for all the settings. Furthermore, there is no evidence of overfitting, since the differences among testing and training for all the scenarios are not big either. Both MSE and IS are quite sensible to reductions of one order of magnitude in the variance of the random spatial effect. The finer the resolution from the data generated the better our downscalling performs, which is an expected result.

Results from Table~\ref{table_time} show that the elapsed time to perform the estimation increases substantially when the resolution is finer. In summary, our downscalling method perform better in terms of fit with finer resolutions, but this improvement is relatively expensive in terms of computational time. 

\begin{table}
\small
\caption{\label{table_time} Elapsed time (in minutes) by scenario, resolution, and model}
\centering
\begin{tabular}[t]{ccccc}
\hline
\multirow{2}{*}{Model} &\multirow{2}{*}{Scenario} & \multicolumn{3}{c}{Resolution} \\
\cline{3-5}
 & & 1 & 2 & 3\\
 \midrule
\multirow{3}{*}{1} & 1 & 0.1309 & 0.2254 & 0.3770\\
 & 2 & 0.1608 & 0.2289 & 0.3759\\
 & 3 & 0.1593 & 0.2279 & 0.4283\\
\midrule
\multirow{3}{*}{2} & 1 & 0.6885 & 1.2371 & 1.6814\\
 & 2 & 0.9042 & 1.2554 & 1.8950\\
 & 3 & 0.8310 & 1.1442 & 2.1358\\
\bottomrule
\end{tabular}
\end{table}

\section{Analysis of NARCCAP Data}\label{sec4}

\begin{figure}[h!]
	\centering
	\includegraphics[scale=0.8]{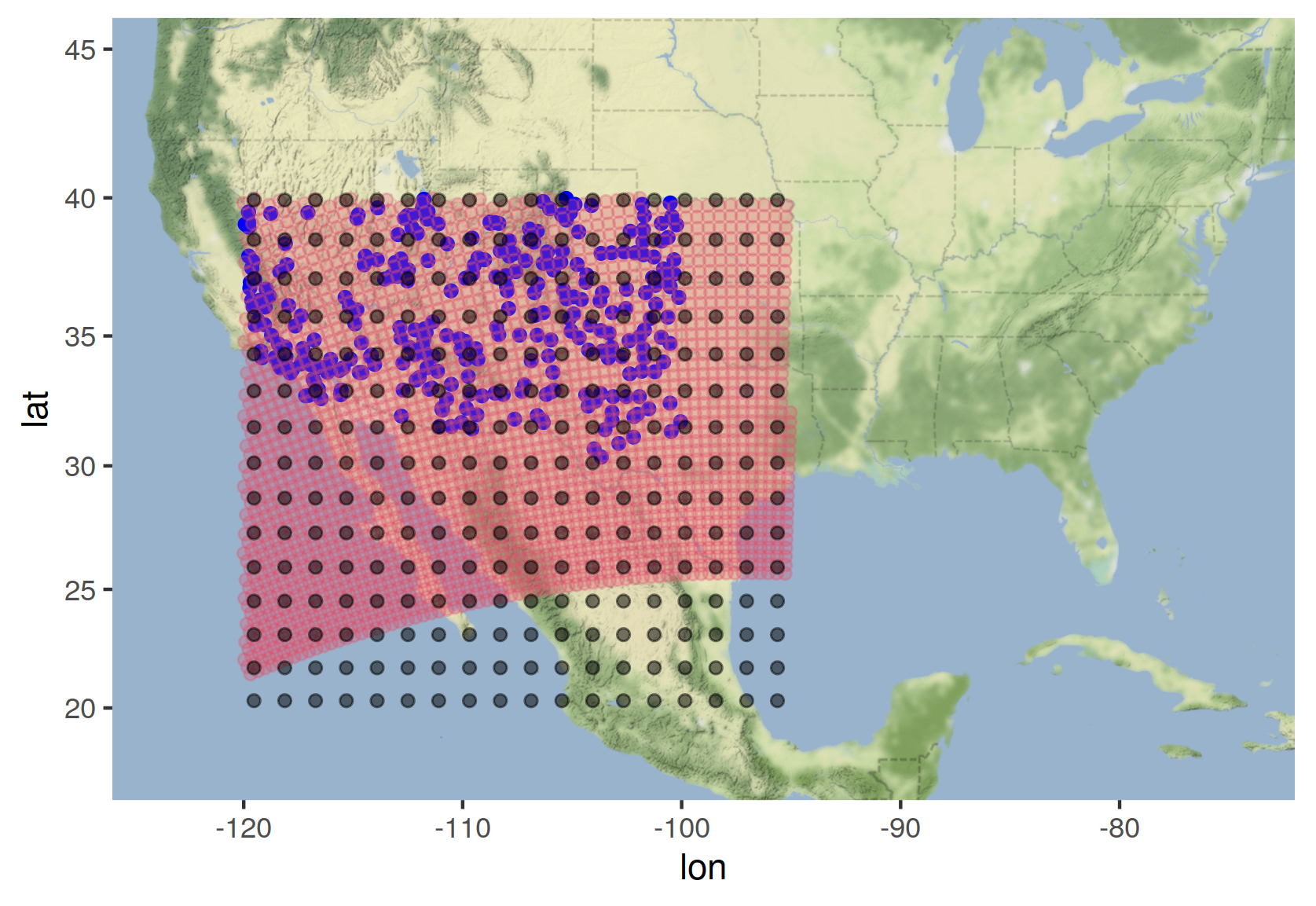}
	\caption{Coordinates for the Canadian Regional Climate Model (CRCM) (red points), the Community Climate System Model (CCSM) in the North American Region (black points) and observed temperatures (blue points).}
	\label{fig1:data}
\end{figure}

Our main source of data is NARCCAP \citep{https://doi.org/10.5065/D6RN35ST}. Our data set consists of mean air surface temperature (ts) measured in Kelvin from the Canadian Regional Climate Model (CRCM) \citep{caya1995description}, seasonal mean pressure vertical velocity measured in Pa/s (OMEGA) and mean air surface temperature (TREFHT) measured in Kelvin from the Community Climate System Model (CCSM), which is the global model embedded in the CRCM \footnote{The complete data sets can be accessed through: \url{https://www.earthsystemgrid.org/project/NARCCAP.html}}. The two climate models have different resolutions and projections, as presented in Figure \ref{fig1:data}, for the study area of interest.

The area of study was defined using the intersection between NARCCAP data and the Monsoon area as defined by \cite{Higgins2006,IPCC2021}, given that it is an important climate region that drives weather changes in Central America and the Caribbean. 

There are 2482 points in the regional-model resolution grid, and 270 points for the global-based one. Monthly data from 1990 to 1998 (inclusive) was used to fit the model, and to test the model we used observed surface air temperature records from the National Climatic Data Center (NCDC) first-order and cooperative observer summary of the day dataset, known as DSI-3200 \cite{NCDC2003}. These data were used and described over the contiguous United States by
\cite{groisman2004contemporary} and in the western states by \cite{Alfaro2006}. The locations of the observed temperatures are shown as blue points in Figure \ref{fig1:data}. The main objective of the emulator is to use the CCSM output to infer the values of the CRCM output. For that propose, it is important to recognize the non-stationary nature of the relationship between the CCSM output and the regional model response, as well as giving the model enough flexibility to describe a non-stationary mean in space and time, together with an acceptable downscaling performance. 

\begin{figure}[htp!]
	\centering
	\includegraphics[scale=0.8]{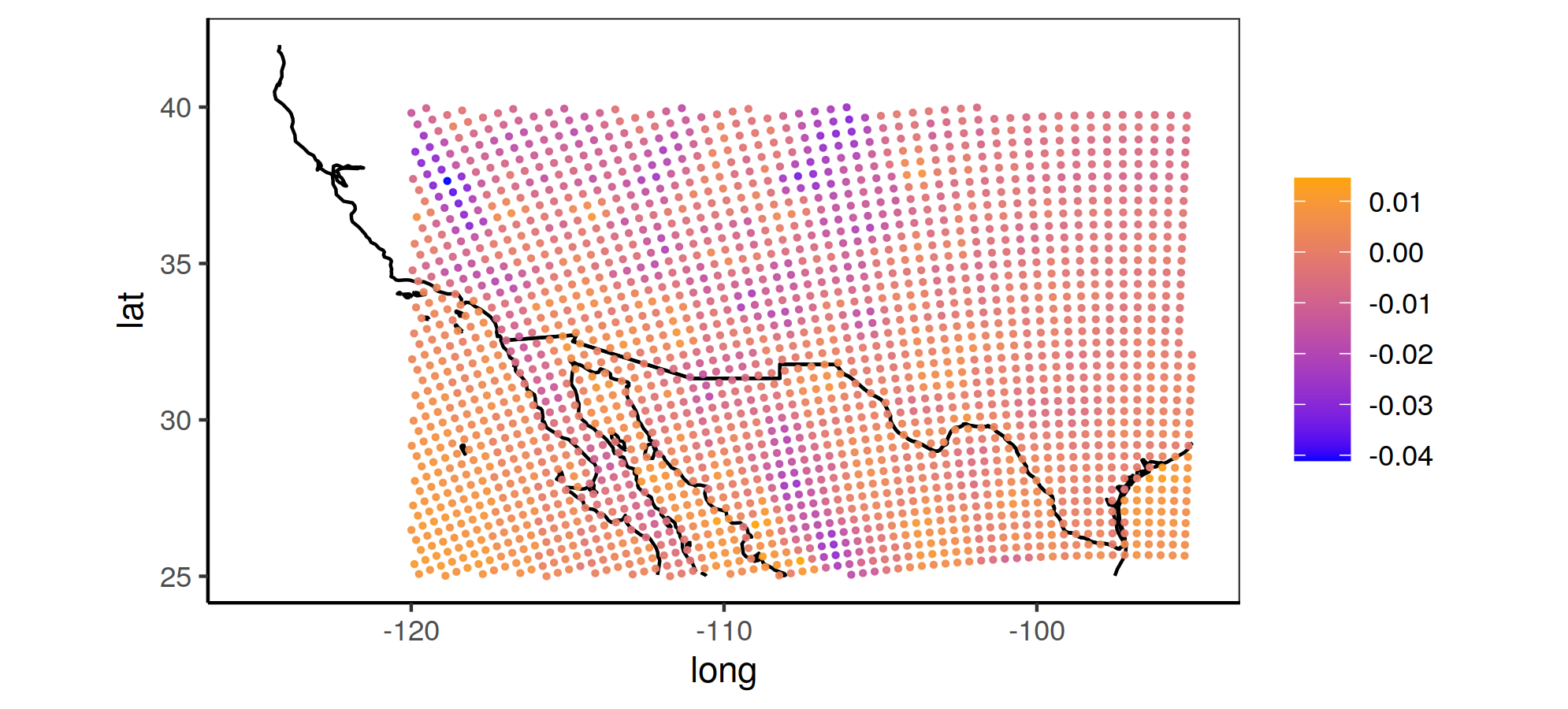}
	\caption{Mean difference between the monthly temperature log-outputs among the regional and global models. The difference was computed over the period 1990-1999 inclusive.}
	\label{fig2:coefficients}
\end{figure}

A descriptive analysis to gather evidence about this non-stationary relationship was done with the data, and it was found that the mean difference \footnote{We computed this mean difference over time, taking both a CRCM grid point and its closest CCSM location.} between both outputs in log-scale was not constant over space (see \ref{fig2:coefficients}), and that defining the mean as constant in space and time could limit the ability of the model to predict the response accurately. A spatio-temporal downscaling emulator that can incorporate spatio-temporal VC is described in the next section.

A formal application of the spatio-temporal downscaling emulator was performed using the NARCCAP data set described in Section 2.  We chose the regional resolution as the main spatial domain of this analysis. The response variable is the temperature from the CRCM Regional Model, measured in Kelvin, for this variable we choose to keep the model units and also to transform the response variable using a log scale, to attain approximately normal behavior. Therefore, we put attention on a varying coefficient model according to \eqref{modelo} with:
\begin{itemize}
	\item $C_t(\cdot) = \log T_t(\cdot)$, where $T_t$ is the temperature and the domains $\mathcal S$ and $\mathcal W$ correspond to the global and regional domains respectively. 
	\item $X_t(\cdot) = OMEGA_t(\cdot)$ Vertical rising rate of air parcels. Provides a measure of large scale rising and sinking motions in the atmosphere. It was obtained from the CCSM model. 
\end{itemize}

The physical reasoning behind the model with the above variables can be justified by the fact that negative values of OMEGA indicate rising and a tendency for convection (and cooler temperatures), and positive values indicate sinking motions associated with warmer temperatures and high pressure (see \citealp{Hostetler2011}). The chosen area of interest is known for presenting this rising motion associated with the monsoonal effect during the boreal summer. Other covariates were considered in the statistical emulator of \eqref{modelo}, such as the sea level pressure (PSL) and horizontal wind components (U and V), but they were not as strongly associated with the log-difference between the regional-level temperature in the CRCM and the global-level temperature in the CCSM.  

In order to compare the predictive performance of the model and choose an adequate emulator for our data, we use the following four alternatives:
\begin{description}
	\item[Model 0:] Non varying intercept model.
	\begin{align*}
		Y_t(w)=\alpha+\gamma_t(w)
	\end{align*}
	\item[Model 1:] Spatio-temporal varying intercept following a random noise process, where the spatial covariance structure follows a Mat\'ern$\left(\nu=1 \right)$.
	\begin{align*}
		\begin{cases}
			Y_t(w)&=\alpha_t(w)+\gamma_t(w)\\
			\alpha_t(w)&= \epsilon_t(w)
		\end{cases}
	\end{align*} 
	\item[Model 2:] Spatio-temporal varying intercept following an $AR(1)$ process, where the spatial covariance structure follows a Mat\'ern$(\nu= 1)$.
	\begin{align*}
		\begin{cases}
			Y_t(w)&=\alpha_t(w)+\gamma_t(w)\\
			\alpha_t(w)&= \rho \alpha_{t-1}(w)+\epsilon_t(w)
		\end{cases}
	\end{align*}

	\item[Model 3:] Spatio-temporal varying intercept following an $AR(1)$ process with a fixed-parameter covariate:
	\begin{align*}
		\begin{cases}
			Y_t(w)&=\alpha_t(w)+\beta \cdot OMEGA_t(w)+\gamma_t(w)\\
			\alpha_t(w)&= \rho \alpha_{t-1}(w)+\epsilon_t(w)
		\end{cases}
	\end{align*}	 
\end{description}

Note that the models are increasing in complexity, from a simple model where the mean behavior does not depend on space or time and its error term is uncorrelated, to a model where the mean structure depends linearly on the external covariate OMEGA and the intercept depends simultaneously on space and time. On the other hand, the error structures of the models also increase in complexity, considering in the last model a slightly more dependent structure than the completely independent case.

The models are fitted using INLA package, in particular by the stochastic partial differential equation (SPDE) approach of \cite{Lindgren2011}. In order to use this method, we train the above models with historical data from 1990 to 1998 (inclusive) at a monthly basis and we use 1999 data as a testing set. The global-model spatial domain is used as knots for the mesh construction in the SPDE approach to reduce the dimensionality of the linear model due to the regional resolution of the RCM. The priors of the non-varying mean coefficients are assumed to be gaussian with the default parameterization of INLA. The prior of the range ($\phi$) and standard deviation ($\sigma$) in each of the Mat\'ern structures is assumed to be a Penalized Complexity (PC) prior (see \citealp{Fuglstad2019}) satisfying: $P[\rho < 700]=0.5$ and $P[\sigma > 0.32] = 0.01$ based on descriptive analysis on the temperature from RCM. A PC prior was also used for the correlation coefficient in the AR1 model, using as hyperparameters 0 and 0.9.

\begin{figure}
	\caption{\label{grafico_temps}Upper panels: Estimated temperatures according to Model 1 (emulator), observed temperatures and RCM. Lower panel: interquantile range for four chosen months during the testing period. Missing values are drawn in gray.}
	
	\centering
	\includegraphics[scale=1]{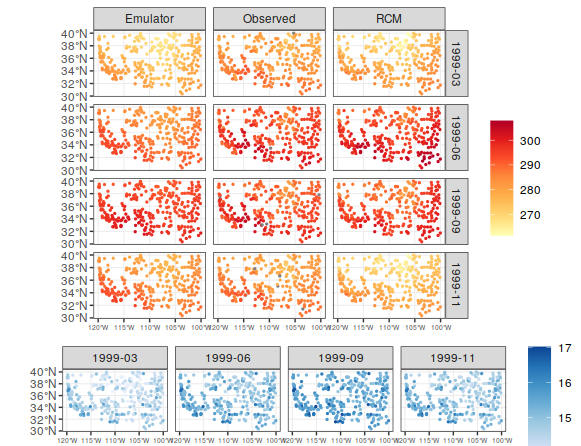}
\end{figure}

In order to compare the predictive performance of the three models, we apply the squared-root MSE and Interval Score between the observed and predicted temperatures over the one-year testing period using the  emulator on each scenario. We chose a one-year testing period to consider the minimum length of time where the annual cycle can be visualized and predicted \cite{Hidalgo2015}. For each location of the observed temperatures (blue dots in Figure \ref{fig1:data}) we computed the metrics with respect to the nearest neighbor belonging to the regional set of locations. Table \ref{tabla_N_uno} contains a comparison of the two predictive metrics for the three models together with the MSE between the observed temperatures and RCM output over the testing period. 
\begin{table}
	\caption{\label{tabla_N_uno}Comparison of models according to the predictive metrics (1990-1998: training period, 1999: testing period) and elapsed time in minutes.}
	\centering
\begin{tabular}{l|l|ll|ll|ll|ll}
	\toprule
	& \multicolumn{1}{c|}{\textbf{RCM}} & \multicolumn{2}{c|}{\textbf{Model 0}} & \multicolumn{2}{c|}{\textbf{Model 1}} & \multicolumn{2}{c|}{\textbf{Model 2}} & \multicolumn{2}{c}{\textbf{Model 3}} \\ \midrule
	Year &  MSE &MSE          & $IS_{.95}$           & MSE          & $IS_{.95}$           & MSE          & $IS_{.95}$           & MSE          & $IS_{.95}$           \\
	\midrule
1990 & 28.96 & 23.23 & 108.48 & 29.21 & 145.11 & 29.01 & 146.20 & 29.02 & 146.22\\
1991 & 32.07 & 18.98 & 96.64 & 32.16 & 153.97 & 31.94 & 155.17 & 31.93 & 155.18\\
1992 & 29.35 & 29.96 & 129.75 & 29.72 & 147.84 & 29.50 & 148.91 & 29.50 & 148.93\\
1993 & 28.28 & 28.87 & 128.90 & 28.77 & 138.62 & 28.55 & 139.78 & 28.56 & 139.82\\
1994 & 28.31 & 29.84 & 127.32 & 28.78 & 141.38 & 28.56 & 142.64 & 28.57 & 142.67\\
1995 & 23.56 & 24.49 & 110.77 & 23.86 & 128.43 & 23.67 & 129.60 & 23.66 & 129.61\\
1996 & 39.79 & 29.87 & 130.64 & 39.98 & 172.70 & 39.77 & 174.02 & 39.78 & 174.04\\
1997 & 23.88 & 20.76 & 97.22 & 24.29 & 128.93 & 24.10 & 129.99 & 24.10 & 129.99\\
1998 & 24.52 & 15.39 & 78.82 & 24.79 & 129.84 & 24.61 & 130.96 & 24.61 & 131.01\\
\midrule
1999 & 29.41 & 38.38 & 149.91 & 37.85 & 36.52 & 40.32 & 40.42 & 40.78 & 41.17\\
	\midrule
	Elapsed Time  & \multicolumn{1}{c|}{-} & \multicolumn{2}{c|}{2.87}    & \multicolumn{2}{c|}{9.69}    & \multicolumn{2}{c|}{36.55}   & \multicolumn{2}{c}{36.57}   \\ \bottomrule
\end{tabular}
\end{table}
and Table \ref{tabla_N_dos} contains the resulting parameters with their 95\%-predictive intervals for the best model according to the metrics (Model 1). 
\begin{table}
	\caption{\label{tabla_N_dos}Parameter estimates and prediction interval for the best scenario (Model 1).}
	\centering
	\begin{tabular}{r|rrr}
		\toprule
		& \textbf{Lower 95\%}  & \textbf{Estimate} & \textbf{Upper 95\%}\\
		\midrule
		$\beta_0$ & -0.00238  & -0.00166  & -0.00094\\
		$\phi$   & 7.2622   &  7.4539   &   7.6463 \\
		$\sigma$ & 0.0152  & 0.0155    &   0.0157 \\
		$\tau$   & 0.00001977 & 0.00001989  & 0.00002001\\
		\bottomrule
	\end{tabular}
\end{table}
Figure \ref{grafico_temps} shows the predicted values (mean) of the temperatures using the best model, the observed temperatures and the estimated temperatures according to the RCM, everything on the observed locations. All those values are shown for three specific months over the testing period. The lower panel the interquantile range on the testing period and the same locations. 	

Note that there is significant difference among the $IS$ obtained over the testing period for Model 0 and Model 1, which is an indicator that the spatial-temporal structure over the intercept is a valid assumption. If we make the temporal structure as AR(1), there is not a gain in terms of predictive performance, but the addition of OMEGA as covariate is not improving the IS metric at all comparing with respect to Model 2. In terms of computational time, the Model 1 offers an interesting alternative. The downscaling capacity of the model can be visualized in Figure \ref{grafico_temps}. It is interesting to note that the general patterns and trends of the observed temperature was predicted successfully by the emulator in most of the months, but when the observed temperature raises during the summer months, the emulator was not able to capture warmer temperatures and it predicts warmer temperatures than the observed when the observed was cooler over certain locations on September, 1999. Moreover, the predicted temperatures of the emulator are closer to the observed ones in November, 1999 than the estimated ones by the RCM. Parameter estimates and prediction intervals are shown for Model 1 in \ref{tabla_N_dos}, where we were able to do quite precise inference in terms of standard errors for all the parameters of interest, but due to the identifiability issues found in simulation II, it would not be surprising to have large bias in the variance parameters.

\section{Conclusions}\label{sec5}

A spatio-temporal downscaling emulator for regional models was proposed using spatial varying coefficients. Among the advantages of the said model, we are able to estimate non-stationary marginal effects, which means that the downscaling function can vary over space. Furthermore, the model has flexibility to estimate the mean of any variable in space and time, as shown on the real data application, although with some restrictions. Finally, that the model has good prediction results for reproducing the annual cycles for one particular year, as shown on both the simulation results, and the analysis of NARCCAP data. This good results in terms of prediction happens in spite of the possible identification issues observed in the same sections. 

In a test simulation, INLA was by far the best model for both the spatial and spatial temporal versions of the proposed model, when it is compared with alternatives as varycoef, under different tapering coefficients. Moreover, INLA was the fastest method for all the cases, and the approximation with best accuracy to estimate the different parameters from the model and the posterior distribution of the response variable. While other programming languages may provide faster computational times, this study is limited to comparing the possibilities achievable through the use of R.

When dealing with this kind of spatio-temporal models, obtaining prior information about $\tau$ and $\sigma$ is not an easy task, due to the fact that both are sources of variability in the model. An initial prior for $\tau$ can be computed using the variograms, but prior for $\sigma$, as the variance of the varying coefficients, is challenging to define. However, by using PC prior this specification can be solved. For future model extensions, different values can be tested in order to avoid identification issues with the model.

The downscaling emulator showed potential for predicting accurately output from RCM using the output from a GCM. Specifically, temperature from the CRCM can be predicted using mean pressure vertical velocity output from CCSM with an out of sample MSE of approximately 34 units for all models proposed, and with a $95\%$ IS that is significantly lower for Models 2 and 3, compared to the other options, as shown in Table 5. Furthermore, when adding the elapsed time to run the models, and given that Model 1 and Model 2 give very similar predictive performance results, then Model 1 is preferred, since it can be run in less than one third of the time that takes to complete Model 2. 

For future work, there are several directions that can be taken. First, more combinations of RCMs and GCMs can be tested, using the data from NARCCAP. In addition, more variables can be used from the GCM output, in order to improve an emulator with the structure of Model 3. Lastly, another programming language such as Python or Julia can be used with the same methods, in order to improve the computational time.

\section{Computational Resources and Code}

All the analysis were run using an HP Proliant DL380 Gen9 server, Intel(R) Xeon(R) CPU E5-2630 v3 @ 2.40GHz, with 16 logical cores, and 128GB RAM. The R scripts to read and wrangle the NARCCAP data, as well to run all of the models described in this paper are available at https://github.com/LEA-UCR/MRA-ST and were run using RStudio Server 2022.02.2 Build 485. Packages and R version are specified in the Readme repository file.

\section{Acknowledgments}
HH and EA are funded through the following Vicerrector\'ia de Investigaci\'on, Universidad de Costa Rica grants: V.I. B0810, C0074, B9454 (supported
by Fondo de Grupos), EC-497 (VarClim, supported by FEES-CONARE) and C0-610 (supported by Fondo de Est\'imulo).

\bibliographystyle{rss}
\bibliography{biblio}%

\begin{thebibliography}{41}
\expandafter\ifx\csname natexlab\endcsname\relax\def\natexlab#1{#1}\fi
\expandafter\ifx\csname url\endcsname\relax
  \def\url#1{\texttt{#1}}\fi
\expandafter\ifx\csname urlprefix\endcsname\relax\def\urlprefix{URL: }\fi

\bibitem[{NCD(2003)}]{NCDC2003}
 (2003) \textit{Data documentation for data set 3200 (DSI-3200): Surface land
  daily cooperative summary of the day}.
\newblock National Climatic Data Center, Asheville, NC.
\newblock
  \urlprefix\url{https://silo.tips/download/national-climatic-data-center-data-documentation-for-data-set-3200-dsi-3200-july}.

\bibitem[{Alfaro et~al.(2006)Alfaro, Gershunov and Cayan}]{Alfaro2006}
Alfaro, E.~J., Gershunov, A. and Cayan, D. (2006) Prediction of summer maximum
  and minimum temperature over the central and western united states: The roles
  of soil moisture and sea surface temperature.
\newblock \textit{Journal of Climate}, \textbf{19}, 1407--1421.
\newblock
  \urlprefix\url{https://journals.ametsoc.org/view/journals/clim/19/8/jcli3665.1.xml}.

\bibitem[{Amador and Alfaro(2009)}]{Amador2009}
Amador, J.~A. and Alfaro, E.~J. (2009) Métodos de redução: aplicações ao
  clima, clima, variabilidade climática e mudanças climáticas.
\newblock \textit{REVIBEC - REVISTA IBEROAMERICANA DE ECONOMÍA ECOLÓGICA},
  39--52.
\newblock
  \urlprefix\url{https://redibec.org/ojs/index.php/revibec/article/view/260}.

\bibitem[{Barboza et~al.(2014)Barboza, Li, Tingley and Viens}]{Barboza2014}
Barboza, L., Li, B., Tingley, M. P.~M. and Viens, F. G.~F. (2014)
  Reconstructing past temperatures from natural proxies and estimated climate
  forcings using short- and long-memory models.
\newblock \textit{The Annals of Applied Statistics}, \textbf{8}, 1966--2001.

\bibitem[{Barboza et~al.(2019)Barboza, Emile-Geay, Li and He}]{Barboza2019}
Barboza, L.~A., Emile-Geay, J., Li, B. and He, W. (2019) Efficient
  reconstructions of common era climate via integrated nested laplace
  approximations.
\newblock \textit{Journal of Agricultural, Biological, and Environmental
  Statistics}, \textbf{24}, 535--554.
\newblock
  \urlprefix\url{https://link.springer.com/article/10.1007/s13253-019-00372-4}.

\bibitem[{Blangiardo et~al.(2013)Blangiardo, Cameletti, Baio and
  Rue}]{Blangiardo2013}
Blangiardo, M., Cameletti, M., Baio, G. and Rue, H. (2013) {Spatial and
  spatio-temporal models with R-INLA}.
\newblock \textit{Spatial and Spatio-temporal Epidemiology}, \textbf{7},
  39--55.
\newblock \urlprefix\url{http://dx.doi.org/10.1016/j.sste.2013.07.003}.

\bibitem[{Bracher et~al.(2021)Bracher, Ray, Gneiting and Reich}]{Bracher2021}
Bracher, J., Ray, E.~L., Gneiting, T. and Reich, N.~G. (2021) Evaluating
  epidemic forecasts in an interval format.
\newblock \textit{PLOS Computational Biology}, \textbf{17}, e1008618.
\newblock
  \urlprefix\url{https://journals.plos.org/ploscompbiol/article?id=10.1371/journal.pcbi.1008618}.

\bibitem[{Castruccio et~al.(2014)Castruccio, McInerney, Stein, {Liu Crouch},
  Jacob and Moyer}]{Castruccio2014}
Castruccio, S., McInerney, D.~J., Stein, M.~L., {Liu Crouch}, F., Jacob, R.~L.
  and Moyer, E.~J. (2014) {Statistical Emulation of Climate Model Projections
  Based on Precomputed GCM Runs*}.
\newblock \textit{Journal of Climate}, \textbf{27}, 1829--1844.
\newblock
  \urlprefix\url{http://journals.ametsoc.org/doi/abs/10.1175/JCLI-D-13-00099.1}.

\bibitem[{Caya et~al.(1995)Caya, Laprise, Giguere, Bergeron, Blanchet, Stocks,
  Boer and McFarlane}]{caya1995description}
Caya, D., Laprise, R., Giguere, M., Bergeron, G., Blanchet, J., Stocks, B.,
  Boer, G. and McFarlane, N. (1995) Description of the canadian regional
  climate model.
\newblock In \textit{Boreal Forests and Global Change}, 477--482. Springer.

\bibitem[{Dambon et~al.(2021)Dambon, Sigrist and Furrer}]{Dambon2021}
Dambon, J.~A., Sigrist, F. and Furrer, R. (2021) {Maximum likelihood estimation
  of spatially varying coefficient models for large data with an application to
  real estate price prediction}.
\newblock \textit{Spatial Statistics}, \textbf{41}, 100470.
\newblock \urlprefix\url{https://doi.org/10.1016/j.spasta.2020.100470}.

\bibitem[{Finley(2011)}]{Finley2011}
Finley, A.~O. (2011) Comparing spatially-varying coefficients models for
  analysis of ecological data with non-stationary and anisotropic residual
  dependence.
\newblock \textit{Methods in Ecology and Evolution}, \textbf{2}, 143--154.
\newblock
  \urlprefix\url{https://onlinelibrary.wiley.com/doi/full/10.1111/j.2041-210X.2010.00060.x
  https://onlinelibrary.wiley.com/doi/abs/10.1111/j.2041-210X.2010.00060.x
  https://besjournals.onlinelibrary.wiley.com/doi/10.1111/j.2041-210X.2010.00060.x}.

\bibitem[{Franco-Villoria et~al.(2019)Franco-Villoria, Ventrucci and
  Rue}]{Franco2019}
Franco-Villoria, M., Ventrucci, M. and Rue, H. (2019) {A unified view on
  Bayesian varying coefficient models}.
\newblock \textit{Electronic Journal of Statistics}, \textbf{13}, 5334 -- 5359.
\newblock \urlprefix\url{https://doi.org/10.1214/19-EJS1653}.

\bibitem[{Fuglstad et~al.(2019)Fuglstad, Simpson, Lindgren and
  Rue}]{Fuglstad2019}
Fuglstad, G.-A., Simpson, D., Lindgren, F. and Rue, H. (2019) {Constructing
  Priors that Penalize the Complexity of Gaussian Random Fields}.
\newblock \textit{Journal of the American Statistical Association},
  \textbf{114}, 445--452.
\newblock
  \urlprefix\url{https://www.tandfonline.com/doi/full/10.1080/01621459.2017.1415907}.

\bibitem[{Gao et~al.(2022)Gao, Director, Bitz and Raftery}]{Gao2022}
Gao, P.~A., Director, H.~M., Bitz, C.~M. and Raftery, A.~E. (2022)
  Probabilistic forecasts of arctic sea ice thickness.
\newblock \textit{Journal of Agricultural, Biological, and Environmental
  Statistics}, \textbf{27}, 280--302.
\newblock
  \urlprefix\url{https://link.springer.com/article/10.1007/s13253-021-00480-0}.

\bibitem[{Gelfand et~al.(2003)Gelfand, Kim, Sirmans and
  Banerjee}]{Gelfand2003a}
Gelfand, A.~E., Kim, H.-J., Sirmans, C.~F. and Banerjee, S. (2003) {Spatial
  Modeling With Spatially Varying Coefficient Processes}.
\newblock \textit{Journal of the American Statistical Association},
  \textbf{98}, 387--396.
\newblock
  \urlprefix\url{http://www.tandfonline.com/doi/abs/10.1198/016214503000170}.

\bibitem[{Gneiting and Raftery(2007)}]{Gneiting2007a}
Gneiting, T. and Raftery, A.~E. (2007) {Strictly Proper Scoring Rules,
  Prediction, and Estimation}.
\newblock \textit{Journal of the American Statistical Association},
  \textbf{102}, 359--378.
\newblock
  \urlprefix\url{http://amstat.tandfonline.com/doi/abs/10.1198/016214506000001437{\#}.UfauueGm1As}.

\bibitem[{Gramacy and Lee(2012)}]{Gramacy2012}
Gramacy, R.~B. and Lee, H.~K. (2012) Adaptive design and analysis of
  supercomputer experiments.
\newblock \textit{https://doi.org/10.1198/TECH.2009.0015}, \textbf{51},
  130--145.

\bibitem[{Groisman et~al.(2004)Groisman, Knight, Karl, Easterling, Sun and
  Lawrimore}]{groisman2004contemporary}
Groisman, P.~Y., Knight, R.~W., Karl, T.~R., Easterling, D.~R., Sun, B. and
  Lawrimore, J.~H. (2004) Contemporary changes of the hydrological cycle over
  the contiguous united states: Trends derived from in situ observations.
\newblock \textit{Journal of hydrometeorology}, \textbf{5}, 64--85.

\bibitem[{Guinness(2022)}]{Guinness2022}
Guinness, J. (2022) Inverses of matérn covariances on grids.
\newblock \textit{Biometrika}, \textbf{109}, 535--541.
\newblock
  \urlprefix\url{https://academic.oup.com/biomet/article/109/2/535/6168989}.

\bibitem[{Heaton et~al.(2018)Heaton, Datta, Finley, Furrer, Guinness,
  Guhaniyogi, Gerber, Gramacy, Hammerling, Katzfuss, Lindgren, Nychka, Sun and
  Zammit-Mangion}]{Heaton2018}
Heaton, M.~J., Datta, A., Finley, A.~O., Furrer, R., Guinness, J., Guhaniyogi,
  R., Gerber, F., Gramacy, R.~B., Hammerling, D., Katzfuss, M., Lindgren, F.,
  Nychka, D.~W., Sun, F. and Zammit-Mangion, A. (2018) A case study competition
  among methods for analyzing large spatial data.
\newblock \textit{Journal of Agricultural, Biological and Environmental
  Statistics}, \textbf{24}, 398--425.
\newblock \urlprefix\url{https://doi.org/10.1007/s13253-018-00348-w}.

\bibitem[{Hernanz et~al.(2022)Hernanz, Garc{\'{i}}a-Valero, Dom{\'{i}}nguez,
  Ramos-Calzado, Pastor-Saavedra and Rodr{\'{i}}guez-Camino}]{Hernanz2022}
Hernanz, A., Garc{\'{i}}a-Valero, J.~A., Dom{\'{i}}nguez, M., Ramos-Calzado,
  P., Pastor-Saavedra, M.~A. and Rodr{\'{i}}guez-Camino, E. (2022) {Evaluation
  of statistical downscaling methods for climate change projections over Spain:
  Present conditions with perfect predictors}.
\newblock \textit{International Journal of Climatology}, \textbf{42}.

\bibitem[{Hidalgo and Alfaro(2015)}]{Hidalgo2015}
Hidalgo, H.~G. and Alfaro, E.~J. (2015) Skill of cmip5 climate models in
  reproducing 20th century basic climate features in central america.
\newblock \textit{International Journal of Climatology}, \textbf{35},
  3397--3421.
\newblock
  \urlprefix\url{https://onlinelibrary.wiley.com/doi/full/10.1002/joc.4216
  https://onlinelibrary.wiley.com/doi/abs/10.1002/joc.4216
  https://rmets.onlinelibrary.wiley.com/doi/10.1002/joc.4216}.

\bibitem[{Higgins et~al.(2006)Higgins, Ahijevych, Amador, Barros, Berbery,
  Caetano, Carbone, Ciesielski, Cifelli, Cortez-Vasquez, Douglas, Douglas,
  Emmanuel, Fairall, Gochis, Gutzler, Jackson, Johnson, King, Lang, Lee,
  Lettenmaier, Lobato, Magaña, Meiten, Mo, Nesbitt, Ocampo-Torres, Pytlak,
  Rogers, Rutledge, Schemm, Schubert, White, Williams, Wood, Zamora and
  Zhang}]{Higgins2006}
Higgins, W., Ahijevych, D., Amador, J., Barros, A., Berbery, E.~H., Caetano,
  E., Carbone, R., Ciesielski, P., Cifelli, R., Cortez-Vasquez, M., Douglas,
  A., Douglas, M., Emmanuel, G., Fairall, C., Gochis, D., Gutzler, D., Jackson,
  T., Johnson, R., King, C., Lang, T., Lee, M.~I., Lettenmaier, D., Lobato, R.,
  Magaña, V., Meiten, J., Mo, K., Nesbitt, S., Ocampo-Torres, F., Pytlak, E.,
  Rogers, P., Rutledge, S., Schemm, J., Schubert, S., White, A., Williams, C.,
  Wood, A., Zamora, R. and Zhang, C. (2006) The name 2004 field campaign and
  modeling strategy.
\newblock \textit{Bulletin of the American Meteorological Society},
  \textbf{87}, 79--94.
\newblock
  \urlprefix\url{https://journals.ametsoc.org/view/journals/bams/87/1/bams-87-1-79.xml}.

\bibitem[{Hostetler et~al.(2011)Hostetler, Alder and Allan}]{Hostetler2011}
Hostetler, S., Alder, J. and Allan, A. (2011) {Dynamically Downscaled Climate
  Simulations over North America: Methods, Evaluation, and Supporting
  Documentation for Users}.
\newblock \textit{Tech. rep.}, : U.S. Geological Survey Open-File Report
  2011-1238.

\bibitem[{IPCC(2021)}]{IPCC2021}
IPCC (2021) Annex {V}: Monsoons.
\newblock In \textit{Climate Change 2021: The Physical Science Basis.
  Contribution of Working Group I to the Sixth Assessment Report of the
  Intergovernmental Panel on Climate Change} (eds. V.~Masson-Delmotte, P.~Zhai,
  A.~Pirani, S.~Connors, C.~Péan, S.~Berger, N.~Caud, Y.~Chen, L.~Goldfarb,
  M.~Gomis, M.~Huang, K.~Leitzell, E.~Lonnoy, J.~Matthews, T.~Maycock,
  T.~Waterfield, O.~Yelekçi, R.~Yu and B.~Zhou), 2193--2204. Cambridge, United
  Kingdom and New York, NY, USA: Cambridge University Press.

\bibitem[{Katzfuss(2017)}]{katzfuss_multi-resolution_2017}
Katzfuss, M. (2017) A {Multi}-{Resolution} {Approximation} for {Massive}
  {Spatial} {Datasets}.
\newblock \textit{Journal of the American Statistical Association},
  \textbf{112}, 201--214.
\newblock \urlprefix\url{https://doi.org/10.1080/01621459.2015.1123632}.

\bibitem[{Katzfuss and Hammerling(2016)}]{Katzfuss2016}
Katzfuss, M. and Hammerling, D. (2016) Parallel inference for massive
  distributed spatial data using low-rank models.
\newblock \textit{Statistics and Computing}, \textbf{27}, 363--375.
\newblock \urlprefix\url{https://doi.org/10.1007/s11222-016-9627-4}.

\bibitem[{Laflamme et~al.(2016)Laflamme, Linder and Pan}]{Laflamme2016}
Laflamme, E.~M., Linder, E. and Pan, Y. (2016) Statistical downscaling of
  regional climate model output to achieve projections of precipitation
  extremes.
\newblock \textit{Weather and Climate Extremes}, \textbf{12}, 15--23.
\newblock \urlprefix\url{https://doi.org/10.1016/j.wace.2015.12.001}.

\bibitem[{Li and Sun(2019)}]{liying2019}
Li, Y. and Sun, Y. (2019) Efficient estimation of nonstationary spatial
  covariance functions with application to high-resolution climate model
  emulation.
\newblock \textit{Statistica Sinica}, \textbf{29}, 1209--1231.

\bibitem[{Lindgren et~al.(2011)Lindgren, Rue and
  Lindstr{\"{o}}m}]{Lindgren2011}
Lindgren, F., Rue, H. and Lindstr{\"{o}}m, J. (2011) {An explicit link between
  Gaussian fields and Gaussian Markov random fields: the stochastic partial
  differential equation approach}.
\newblock \textit{Journal of the Royal Statistical Society: Series B
  (Statistical Methodology)}, \textbf{73}, 423--498.
\newblock
  \urlprefix\url{http://doi.wiley.com/10.1111/j.1467-9868.2011.00777.x}.

\bibitem[{Mearns et~al.(2007)Mearns, McGinnis, Arritt, Biner, Duffy, Gutowski,
  Held, Jones, Leung, Nunes, Snyder, Caya, Correia, Flory, Herzmann, Laprise,
  Moufouma-Okia, Takle, Teng, Thompson, Tucker, Wyman, Anitha, Buja, Macintosh,
  McDaniel, O{'}Brien, Qian, Sloan, Strand and
  Zoellick}]{https://doi.org/10.5065/D6RN35ST}
Mearns, L., McGinnis, S., Arritt, R., Biner, S., Duffy, P., Gutowski, W., Held,
  I., Jones, R., Leung, R., Nunes, A., Snyder, M., Caya, D., Correia, J.,
  Flory, D., Herzmann, D., Laprise, R., Moufouma-Okia, W., Takle, G., Teng, H.,
  Thompson, J., Tucker, S., Wyman, B., Anitha, A., Buja, L., Macintosh, C.,
  McDaniel, L., O{'}Brien, T., Qian, Y., Sloan, L., Strand, G. and Zoellick, C.
  (2007) North american regional climate change assessment program dataset.
\newblock \urlprefix\url{https://doi.org/10.5065/D6RN35ST}.

\bibitem[{Mearns et~al.(1999)Mearns, Bogardi, Giorgi, Matyasovszky and
  Palecki}]{Mearns1999}
Mearns, L.~O., Bogardi, I., Giorgi, F., Matyasovszky, I. and Palecki, M. (1999)
  Comparison of climate change scenarios generated from regional climate model
  experiments and statistical downscaling.
\newblock \textit{Journal of Geophysical Research: Atmospheres}, \textbf{104},
  6603--6621.
\newblock \urlprefix\url{https://doi.org/10.1029/1998jd200042}.

\bibitem[{Mearns et~al.(2009)Mearns, Gutowski, Jones, Leung, McGinnis, Nunes
  and Qian}]{Mearns2009}
Mearns, L.~O., Gutowski, W., Jones, R., Leung, R., McGinnis, S., Nunes, A. and
  Qian, Y. (2009) {A Regional Climate Change Assessment Program for North
  America}.
\newblock \textit{Eos, Transactions American Geophysical Union}, \textbf{90},
  311--311.
\newblock \urlprefix\url{http://doi.wiley.com/10.1029/2009EO360002}.

\bibitem[{O'Hagan(2006)}]{OHagan2006}
O'Hagan, A. (2006) {Bayesian analysis of computer code outputs: A tutorial}.
\newblock \textit{Reliability Engineering {\&} System Safety}, \textbf{91},
  1290--1300.
\newblock
  \urlprefix\url{http://linkinghub.elsevier.com/retrieve/pii/S0951832005002383}.

\bibitem[{Overstall and Woods(2016)}]{overstall2016multivariate}
Overstall, A.~M. and Woods, D.~C. (2016) Multivariate emulation of computer
  simulators: model selection and diagnostics with application to a
  humanitarian relief model.
\newblock \textit{Journal of the Royal Statistical Society: Series C (Applied
  Statistics)}.

\bibitem[{Rue et~al.(2009)Rue, Martino and Chopin}]{Rue2009}
Rue, H., Martino, S. and Chopin, N. (2009) {Approximate Bayesian Inference for
  Latent Gaussian Models by Using Integrated Nested Laplace Approximations}.
\newblock \textit{Journal of the Royal Statistical Society . Series B (
  Methodological )}, \textbf{71}, 319--392.

\bibitem[{Sacks et~al.(1989)Sacks, Welch, Mitchell and Wynn}]{Sacks1989}
Sacks, J., Welch, W.~J., Mitchell, T.~J. and Wynn, H.~P. (1989) Design and
  analysis of computer experiments.
\newblock \textit{https://doi.org/10.1214/ss/1177012413}, \textbf{4}, 409--423.
\newblock
  \urlprefix\url{https://projecteuclid.org/journals/statistical-science/volume-4/issue-4/Design-and-Analysis-of-Computer-Experiments/10.1214/ss/1177012413.full
  https://projecteuclid.org/journals/statistical-science/volume-4/issue-4/Design-and-Analysis-of-Computer-Experiments/10.1214/ss/1177012413.short}.

\bibitem[{Wang et~al.(2022)Wang, Mou and Liu}]{Wang2022}
Wang, D., Mou, X. and Liu, Y. (2022) Varying-coefficient regression analysis
  for pooled biomonitoring.
\newblock \textit{Biometrics}, \textbf{78}, 1328--1341.
\newblock \urlprefix\url{https://pubmed.ncbi.nlm.nih.gov/34190334/}.

\bibitem[{Wang et~al.(2004)Wang, Leung, L.~McGregor, Lee, Wang, Ding and
  Kimura}]{wang04}
Wang, Y., Leung, L., L.~McGregor, J., Lee, D.-K., Wang, W.-C., Ding, Y. and
  Kimura, F. (2004) Regional climate modeling: Progress, challenges, and
  prospects.
\newblock \textit{Journal of The Meteorological Society of Japan - J METEOROL
  SOC JPN}, \textbf{82}, 1599--1628.

\bibitem[{Wilby and Wigley(1997)}]{Wilby1997}
Wilby, R. and Wigley, T. (1997) {Downscaling general circulation model output:
  a review of methods and limitations}.
\newblock \textit{Progress in Physical Geography}, \textbf{21}, 530--548.
\newblock
  \urlprefix\url{http://ppg.sagepub.com/cgi/doi/10.1177/030913339702100403}.

\bibitem[{Wood et~al.(2004)Wood, Leung, Sridhar and Lettenmaier}]{Wood2004}
Wood, A.~W., Leung, L.~R., Sridhar, V. and Lettenmaier, D.~P. (2004) Hydrologic
  implications of dynamical and statistical approaches to downscaling climate
  model outputs.
\newblock \textit{Climatic Change}, \textbf{62}, 189--216.
\newblock \urlprefix\url{https://doi.org/10.1023/b:clim.0000013685.99609.9e}.

\end{thebibliography}

\appendix

\section{Appendix: Approximation methods comparison in the spatial case}\label{AppendixA}

We study the fitting performance of two approximation methods for a spatially VC model: Dambon Approach and INLA. For the Dambon Approach, we used the package \textit{varycoef} with tapering of $0.05$ and $0.1$, available on CRAN, and adapted its likelihood evaluations to use it with a MCMC. Finally, we used the INLA package available on CRAN.

Our main focus is to describe how well we can estimate a simulated response according to \eqref{Regional_eq}, and compare goodness-of-fit performance for different structures of spatial correlation when estimating the response variable and the parameters involved, according to elapsed time and the final model estimation in \eqref{modelo}. For that, we use the following statistics: Interval Score (IS) and MSE for each estimated model; where the former metric is defined in equation \eqref{eq:IS} and we complement the performance of each approximation technique by means of the mean elapsed time to estimate the model \eqref{modelo} (in minutes).

In the spatial region $\left[0,1\right]^2$, we set three different resolutions for the regularly-spaced set $\mathcal S$ and the finer set $\mathcal W$. We use $25 \times 10$, $40 \times 20$ and $55 \times 25$ as three different resolution scenarios for this simulation and for each scenario we generate 10 replications, using the notation defined in section \ref{sec3}.

At these locations, we simulate two models according to \eqref{Global_eq} and \eqref{Regional_eq} with $q=1$, and $\alpha=5.6$, $\beta_0=-0.05$, $\beta=0.015$ and  $\beta_1=-0.005$, $\epsilon_t(s)\stackrel{i.i.d}{\sim} N(0,\zeta^2)$ with $\zeta^2=2$ and $\tau^2=1$. In this simulation, we consider a spatially varying intercept with a spatial covariance structure following either a Mat\'ern $(\phi=0.1,\sigma=0.001,\nu=0.8)$ or an Exponential$(\phi=0.1,\sigma=0.001)$. 

\subsection{Simulation Results}

Data is analyzed using a PC prior in all cases according to \cite{Franco2019}. Table~\ref{tab:simulation_prediction} presents the goodness of fit metrics by model and resolution, including the elapsed time in minutes for each of the approximation methods.

\begin{table}[H]	\caption{\label{tab:simulation_prediction}Goodness of fit, and elapsed time (in minutes) by approximation method, covariance structure of the data, and resolution.}
	\centering
	{\scriptsize
		\begin{tabular}{ll|rrp{1cm}|rrp{1cm}|rrp{1cm}}
			\toprule
			\multicolumn{2}{c}{} & \multicolumn{9}{c}{\textbf{Resolution}} \\
			\cmidrule{3-11}
			\multicolumn{2}{c}{ } & \multicolumn{3}{|c|}{$25\times 10$} & \multicolumn{3}{c|}{$40\times 20$} & \multicolumn{3}{c}{$55\times 25$} \\
			\midrule
			Covariance & Model & MSE & $IS_{.95}$ & Elapsed time & MSE & $IS_{.95}$ & Elapsed time & MSE & $IS_{.95}$ & Elapsed time \\
			\midrule
			\multirow{3}{*}{Exponential} & varycoef$_{0.05}$ & 12.328 &  10.176 & 4.52  & 20.598 & 10.579 &  50.713 & 28.517 & 10.536 & 322.653\\
			 & varycoef$_{0.1}$ &12.588   &10.276   &5.80    &20.674  &10.450  &63.44   &28.984&10.593 &450.32 \\
			& INLA     &5e-08	 &0.0837   &8.70 	   &2e-08  &0.0852   &10.44    &1e-08 &0.0847 &17.10  \\
			\midrule
			\multirow{2}{*}{Mat\'ern} & varycoef$_{0.05}$ & 12.309 & 10.176  & 4.55 & 20.339 & 10.553 &  52.464 & 28.858 & 10.484 & 324.973 \\ 
			 & varycoef$_{0.1}$ &12.660	   &10.224   & 5.94  & 20.620	 &10.414  &62.16  &28.963&10.586&430.62  \\ 
			& INLA     &6e-08	&0.0820   & 8.63   &1e-08	  &0.0853 &11.11  &1e-08	&0.0844& 15.81 \\
			\bottomrule
		\end{tabular}
	}
\end{table}

With a relatively small number of points ($40 \times 20$), varycoef has less elapsed time than INLA in both covariance structures, but their predictive metrics are larger. In general, INLA is a very superior option in both time and accuracy, as shown in Table~\ref{tab:simulation_prediction}. It is worth noting that despite choosing two different scenarios for the tapering parameter, both resulted in prediction metrics that differ from those offered by INLA. Furthermore, it is possible to infer that even when using a hypothetical scenario that is "close to diagonal" in the covariance matrix, the values obtained by INLA may not be reached.

Although this article aims to measure the predictive capacity of the emulator proposed in \eqref{modelo}, we found that the range and variance parameters under the Matern distribution has large bias when comparing the parameter estimates with their actual values. As our estimation using INLA uses the SPDE approach on a regular grid, we believe that this behavior may confirm the results obtained in \cite{Guinness2022} where the range parameter is affected when considering higher frequency data under regular grid structures. But as the same author confirms, this possibly may not be a problem from a predictive standpoint. It is important to add that the parameter estimation did not present much of a difference for the approximation methods, in terms of bias or variability. 

Given these results, we select INLA as the approximation method of choice for sections \ref{sec3} and \ref{sec4}.

\end{document}